\documentclass [12pt]{article}
\usepackage {amssymb}
\usepackage {amsmath}
\usepackage{graphicx}
 \usepackage {longtable}

\begin{document}
\begin{center}
\textbf{Scintillator-photodiode type detectors for multi-energy
scanning introscopy}
\end{center}

\bigskip

\begin{center}
V.D.~Ryzhikov, S.V.~Naydenov, D.N.~Kozin, \\ E.K.~Lisetskaya,
A.D.~Opolonin, V.M.~Svishch, T.V.~Kulik
\end{center}

\begin{center}
Concern "Institute for Single Crystals"\, of the National Academy
of Sciences of Ukraine
\end{center}

\bigskip

Results of experimental studies of detector arrays S-PD
(scintillator-photodiode) and PRD (scintillator-photoreceiving
device) used for X-ray digital radiography have shown that there
exist further possibilities to increase spatial resolution of this
system up to 2-3 line pairs per $mm$. Theoretical analysis and
experimental studies show that the two-energy detection method not
only allows one to detect organics on the background of metal, but
also substantially increases (by 3-5 times) the detection ability
of the system as a whole, especially if parameters of the S-PD
pair are optimized, in particular, when ZnSe(Te) is used in the
low-energy circuit. A possibility to distinguish, in principle,
between substances with insignificant differences in atomic number
has been theoretically proven -- by transition to multi-energy
radiography. 3D-imaging has been realized using S-PD detector
arrays.

\bigskip

\section{Introduction}

X-ray digital radiography is a rapidly expanding and one of the
most important methods of modern non-destructive testing [1]. In
this method, alongside with the use of luminescent screens with
subsequent transformation of the image onto SCS-matrices (spatial
charged systems), one of the main technical solutions is
conversion of the penetrating X-ray radiation by the detector
array of  "scintillator-photodiode"\, (S-PD) type with its
subsequent amplification and digitalization of the signal.

Advantages of SCS-devices are instant imaging of all the object
and high spatial resolution (3-5 line pairs per mm). Their
disadvantage is a limited energy range, consequently, limited
steel thickness of the inspected object, as well as higher costs,
as compared with S-PD arrays.

In non-destructive testing systems using S-PD arrays it is
possible to use scintillators of different atomic number, density
and element length, which allows working in the energy range from
20 $keV$ to 16 $MeV$, i.e., steel equivalent thickness is from 100
$\mu m$ to 300 $mm$. The use of two-energy detection systems
solves the problem of distinguishing between substances of similar
density, but different effective atomic numbers. Both these
qualities are not attainable for SCS-matrices.

Our task in developing this method consisted in the maximum use of its
advantages. Specifically, we aimed at increased sensitivity and detecting
ability due to optimization of parameters of the S-PD pair and an extensive
use of the features of two-energy radiography. Transition to multi-energy
radiography was envisaged for detection of substances with close values of
the effective atomic number. The resolution was to be increased due to
modernization of the design and making smaller the detector aperture. And,
finally, passing from two- to three-dimensional imaging was also essential.

The above-listed directions in the system improvement were the aims of our
studies in this work.

\section{Experimental procedures}

Measurements of the detector sensitivity and light output of the
crystals were carried out on a testing board using X-ray sources
IRI ($U_{a}=40-200 kV$, $I_{a}=0.4-1.0 mA$, W-anode) and REIS
($U_{a}=5-45 kV$, $I_{a}=5-50 \mu A$, Ag-anode) and an optical
power meter "Kvarts-01". Time characteristics were measured
using a testing board designed for afterglow measurements [2].

To obtain shadow X-ray images, we used testing boards "Poliscan"
(a 128-channel array of photodiode-based detectors) and
``Photocell'' (for 32-, 64-, 128- and 1024-channel detectors based
on PRD). Using the Photocell board, sixty 2D-images of a small
object were obtained at different angles (with step 6), from which
a 3D image was reconstructed.

For detection of X-ray radiation, we used detector arrays of types
S-PD and PRD. Detectors of PRD type includes an array of
photodiodes (32, 64, 128 and 1024 channels), amplifier and
commutator mounted on one silicon crystal.

We used standard scintillation crystals CsI(Tl), CdWO$_{4}$ and an
original scintillator ZnSe(Te) developed by STC "Institute for
Single Crystals" [2]. The photodiodes used were obtained from
producers CCB Ritm, SPO Bit, Ukraine, and Hamamatsu, as well as
PRD from SPO Bit.

Main parameters of scintillation materials for S-PD and S-PRD detectors are
presented in Table 1.

\bigskip

\begin{table}
\begin{tabular}
{|p{160pt}|p{50pt}|p{50pt}|p{50pt}|}

\hline Parameter & ZnSe(Te)& CsI(Tl)& CdWO$_{4}$ \\

\hline Conversion efficiency& 19,4& 15& 3,5
\\ \hline Decay time $\tau $, $\mu s$& 5-7& 1& 5-7 \\

\hline Density, $g/cm^{3}$& 5,4& 4,5& 7,9 \\

\hline Effective atomic number, $Z_{eff}$& 33& 52& 66 \\

\hline Luminescence maximum $\lambda _{m}$ at 300 $K$, $nm$ & 610
& 570 & 490 \\

\hline Afterglow, \% (after 10 $\mu s$)& 0,05& 1-8& 0,05 \\

\hline Absorption coefficient at $\lambda _{m}$, $cm^{-1}$ &
0,1-0,3& 0,05 & 0,02-0,05 \\

\hline Refraction coeff. at $\lambda _{m}$ & 2,4& 1,79& 2,25
\\

\hline Total internal reflection angle $\gamma $ for  \par
$n_{1}$=1,4 \par $n_{2}$=1,5& 37\r{}  \par 40\r{} & 57,17\r{}
\par 63,25\r{} & 42,75\r{}  \par 16,45\r{} $^{}$ \\ \hline
Spectral matching coefficient with Si-PD, $K_{SC}$& 0,49& 0,34&
0,27 \\

\hline Radiation stability limit, $rad$& 10$^{8}$ & 10$^{4}$&
10$^{7}$ \\

\hline Energy of K-jumps, $keV$ & 9,7 \par 12,6& 33,2 \par 35,9&
26,1 \par 69,5 \\ \hline
\end{tabular}
\caption{Characteristics of scintillators used in
"scintillator-photodiode" system.}
\end{table}

\section{Theoretic analyze of multi-energy radiography}

In the recent years, an important role in the modern digital
radiography has been taken by the so-called multi-energy method
[1-3]. Its application in X-ray introscopy systems [4-5] is
promoted by new inspection possibilities that are provided by this
method as compared with conventional radiography means. Thus, two-
and three-energy introscopy systems can not only distinguish
between organics and inorganics, but also discern one organic
compound from another.

A physical model of the multi-energy radiography is based on the
property of exponential attenuation of hard ionizing radiation in
the inspected object. An important element of multi-energy systems
is digital processing of data arrays that are coming
simultaneously from detectors of different types. The computer
does also allow presentation of the output signal in the most
convenient form (logarithmic and normalized). Theoretically,
multi-energy radiography is described by a system of linear
equations [3]
\begin{equation}\label{eq1}
R_{\gamma}  = {\sum\limits_{\delta = 1}^{M} {M_{\gamma
{\kern 1pt} \delta} }} X_{\delta}  ;\quad M_{\gamma {\kern 1pt}
\delta}  = \hat {M}\left( {E_{\gamma};\, E_{\delta} }  \right)
\end{equation}
where $R_{\gamma}=R\left(E_{\gamma}\right)=\ln \left[ {F_{0}\left(
E_{\gamma}\right)}\left/ {F \left(E_{\gamma}\right)} \right.
\right]$ is the system reflex; $F_{0}$ and $F$ -- output signals
from the background (in the absence of an object) and from the
object $E_{\gamma}$ one of the selected radiation energies; $M$ --
the order of multi-energeticity. For two-energy radiography,
$\gamma = 1,2$; $\delta = 1,2$ and $E_{\gamma}  = {\left\{ {E_{1}
;E_{2}} \right\}}$. The unknown $X$ correspond to physical
parameters under control. The monitoring matrix $\hat {M}$ does
not depend upon properties of the studied object. Its components
are determined after calibration measurements on samples of the
known composition (effective atomic number $Z_{\delta}$ and
density $\rho _{\delta}$) and thickness $L_{\delta}$: $C_{\gamma
{\kern 1pt} \delta} = R\left( {E_{\gamma};\, Z_{\delta} ,\rho
_{\delta} ,L_{\delta} } \right)$.

In the two-energy radiography, the effective atomic number of an
unknown material can be reconstructed using the formula obtained
from the general equations system (\ref{eq1})
\begin{equation}\label{eq2}
Z_{eff} = Z\left( Y \right) = Z_{1} \left[ \frac{Y - \Delta
_{1}}{Y - \Delta _{2}} \right]^{1/3}; \quad Y = \frac{C_{12}R_{2}
- C_{22}R_{1}} {C_{11}R_{2} - C_{21}R_{1}}
\end{equation}
Here auxiliary values are introduced: $\Delta _{1} = \left( Z_{2}
\left/ Z_{1} \right. \right)^{3} \left(\rho _{2}L_{2} \left/ \rho
_{1}L_{1} \right. \right)$; $\Delta _{2} = \Delta _{1} \left(
Z_{1} \left/ Z_{2} \right. \right)^{2}$; $Y$~is the reduced
(relative) output signal. The expressions obtained depend only on
radiographic measurement data $R_{1,2} = R\left( {E_{1,2}}
\right)$ from the detector pair and the calibration (reference)
data $C_{11} ,C_{12} ,C_{21} ,C_{22} $ and $\Delta _{1} ,\Delta
_{2} $. It should be stressed that this is a direct method for
determination of the effective atomic number.

We have to note that all the existing developments in the field of
two- and even three-energy radiography involve only qualitative
methods of effective atomic number determination. Their accuracy
does not exceed 50\% for compounds with large $Z_{eff} $. Within
these limits, it is possible only to distinguish between organics
and inorganics, or heavy alloys (of iron, $Z_{eff} \approx 26$)
from light alloys (of aluminum, $Z_{eff} \approx 13$). However,
possibilities of the two-energy radiography have not yet been
fully studied and used.

Our analysis have shown that the effective atomic number of the
material can be reconstructed with the accuracy of 5-10\% using
the formula (3) (see Fig.1a). This is by an order of magnitude
better than the highest level obtained by now. Obviously, such an
improvement in diagnostics will allow one to identify inorganic
and inorganic materials with much higher probability (up to
80-90\% ). This increase in sensitivity of effective atomic number
determination is sufficient to allow detection of, say, explosives
on the background of organic substances (and not just inorganics).
Among practical application of this method is detection of plastic
explosives inside postal parcels and letters, transported goods
and merchandise, etc.

Thus, remaining within the limits of two-energy radiography, one
can carry out quantitative diagnostics of the effective atomic
number and density of materials with accuracy of 5-10\% . There
will be two controlling parameters, not one, which also
contributes to the inspection efficiency. Moreover, transition
from two- to three-energy radiography ensures better control not
only of the effective atomic number, but also of atomic (molar)
concentrations of a complex chemical compound. The number of
reconstructed parameters corresponds to the order of
multi-energeticity. In the case of organic compounds, it is
possible to determine the relative content of two or three
``heaviest'' simple elements -- carbon ($Z = 6$), nitrogen ($Z =
7$) and oxygen ($Z = 8$). As a result, there is a high probability
to clearly distinguish between an explosive (with higher content
of the latter two elements [7]) and other organic compounds.
Combination of diagnostics involving atomic numbers and that
involving concentrations of predominant atomic components seems to
be a key factor in the solution of this important problem. It is
essential that practical realization of the outlined approach is
based upon the use of mobile multi-energy X-ray inspection systems
for non-destructive testing, and not of large-sized and expensive
installations of neutron technologies.

In X-ray customs inspection systems, distinction between organic
and inorganic substances is made by the two-energy method, which
consists in the following. The X-ray tube has a continuous
spectrum (Fig.1b). Passing through the inspected object, the
initial flux of X-ray radiation is partially absorbed, and in the
luminescence spectrum the ratio of low-energy and high-energy
contributions is changed. The higher is the effective atomic
number $Z_{eff}$ of a substance, the smaller is fraction of the
low-energy radiation. To determine the ratio between high- and
low-energy parts of the spectrum, a "two-energy"\, detecting array
is used, which consists of two detector arrays of S-PD type placed
one after another. The array that is closer to the X-ray source is
optimized for detection of low-energy (30-50 $keV$) radiation, and
the second array, which is placed behind the first, is optimized
for detection of high-energy (80-120 $keV$) radiation. The total
signal from the low- and high-energy detectors is characterized by
the product $\rho \,l$, where $\rho $ is density of the inspected
object, $l$ is its radiation length. The ratio of these detector
signals depends upon $Z_{eff}$.

\begin{figure}[ht]
\begin{center}
\textbf{(a)}
\includegraphics*[scale=1.0]{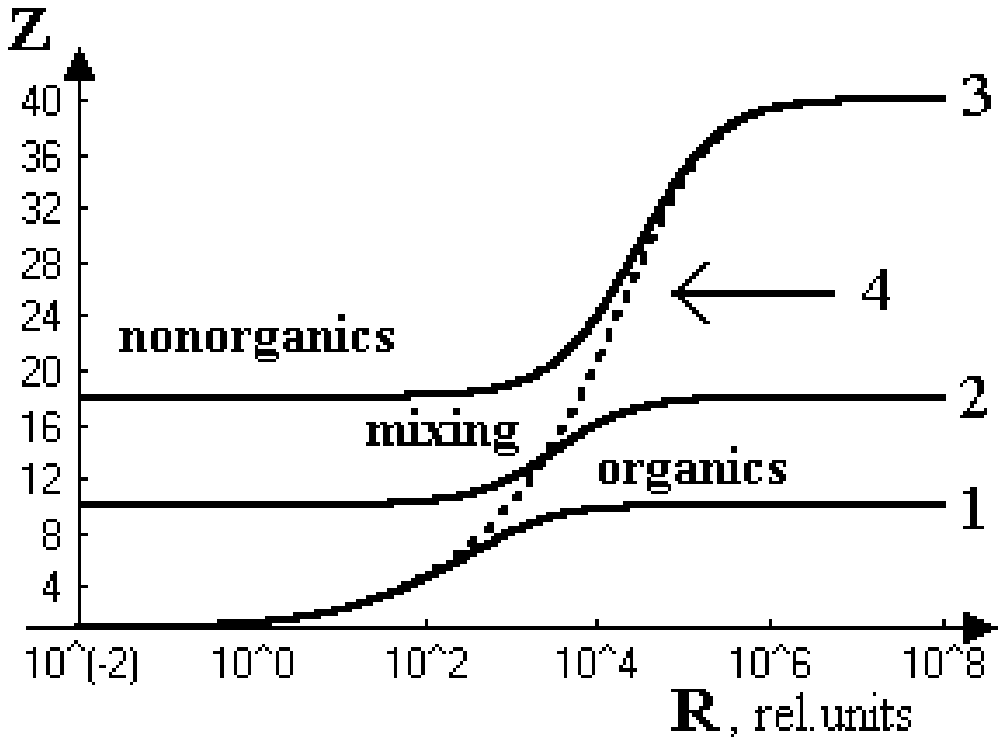}
\end{center}
\begin{center}
\textbf{(b)}
\includegraphics*[scale=0.8]{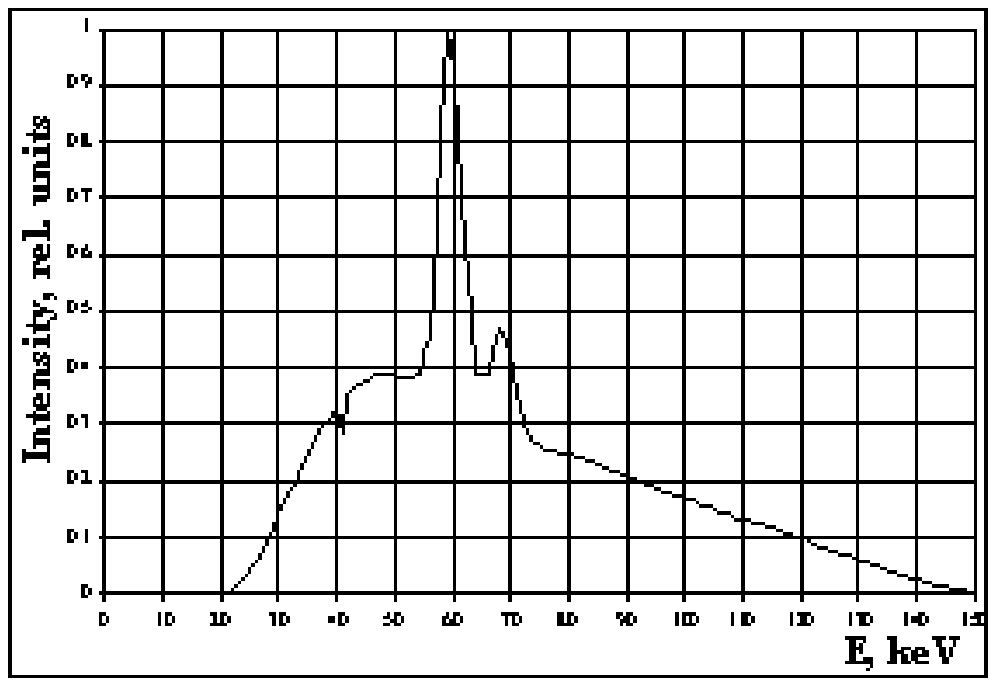}
\end{center}
\end{figure}

\begin{figure}[t]
\begin{center}
\textbf{(c)}
\includegraphics*[scale=1.0]{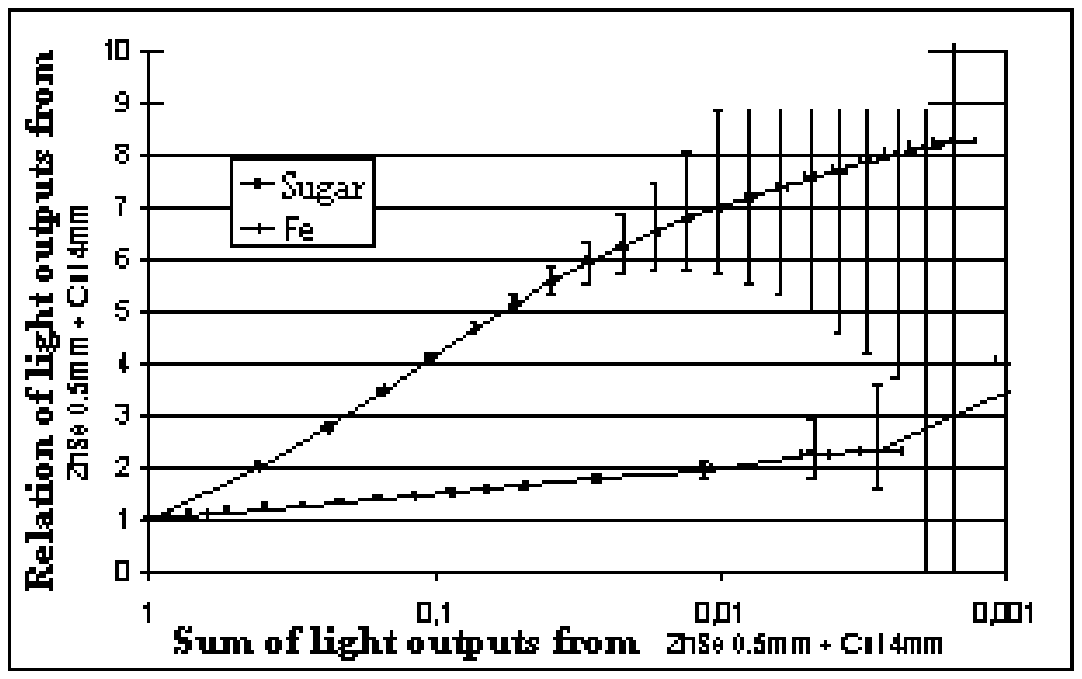}
\end{center}
\caption{Two-energy method for determination of effective atomic
number $Z_{eff}$. (a) Theoretical plots of the effective atomic
number $Z\left(R\right)$ as function of the ratio $R = R_{1}
\left/R_{2}\right.$ of two output signals in the two-energy
radiography inspection of $Z_{eff} $ in different ranges: 1 --
$\left[Z_{1};\, Z_{2} \right] = \left[10;\,1 \right]$; 2 --
$\left[18;\,10 \right]$; 3 -- $\left[40;\,18 \right]$; 4 --
$\left[40;\,1 \right]$. (b) Spectrum of X-ray tube with W-anode at
150 $kV$. (c) Calculated dependence of $R=R_{1}\left/R_{2}\right.$
ratio on thickness of the studied object (for sugar and Fe )
accounting for quantum noises of the analog-to-digit converter. }
\end{figure}

We have carried out theoretical calculations for the following
model of the two-energy system. As initial radiation spectrum, we
took the spectrum of an X-ray tube at constant anode voltage (150
$kV$); the anode material is tungsten (Fig.1b). As inspected
object, we selected sugar (C$_{17}$H$_{22}$O$_{11}$ -- organics)
and iron (Fe -- inorganics). Two-energy detector arrays include
two types of scintillators: the first one (closer to the radiation
source) -- ZnSe(Te) of thickness 0.5 $mm$, and the second one
(behind the first) -- CsI(Tl) of thickness 4 $mm$. Fluxes of
optical quanta emerging in the scintillators under X-ray
irradiation are considered as output signals. The calculated
signal was normalized and conditionally transformed by a 12-digit
analog-to-digit converter. Confidence intervals related to the
quantum noises were calculated. The results are presented in
Fig.1c.

\section{Spatial resolution and detector sensitivity}

Transition from the film radiography to the digital radiography
has already resulted in important improvements in many fields as
for productivity and quality of X-ray non-destructive testing.
However, medical and non-destructive testing applications
requiring detection of non-homogeneities of less than 0.5 $mm$
size are related to the need to improve spatial resolution at
least to 5-8 line pairs per $mm$ .

Among principal ways to solve this problem, one could note -- 1) development
of multi-channel detector arrays with small step and low aperture; 2) using
principles of geometrical optics to obtain expanded shadow images of the
inspected object; 3) the use of various methods of mathematical processing
for image analysis.

The first method is the most preferable, as it allows one to
obtain a shadow image of the object with the least possible
geometrical distortions. However, this way to improve the
resolution encounters significant technological and technical
difficulties. Our experience in developing S-PD and S-PRD
detectors, as well as detector arrays based thereon, have shown
that development and production of multi-channel diode arrays with
step $ \sim  25 \mu $ (up to one thousand channels on one silicon
crystal) does not evoke major technological problems. However,
compact placement of a large number of pre-amplifiers close to the
diode array (a separate amplifier is required for each photodiode)
becomes problematic already at the detector step of 0.8-0.4 $mm$ .
A solution to this problem could be found by placing PD,
pre-amplifiers and the commutator upon one silicon crystal. On the
basis of such photoreceiving devices (PRD), we have developed
multi-channel detecting modules. A multi-channel detection module
is a multi-channel (16,32, 64, 1024 channels) photoreceiver (PD or
PRD), upon the photosensitive area of which a scintillator plate
is attached by means of an optical adhesive [6].

Scintillator plates of the following types were tested:
\begin{itemize}
\item Plates of discrete scintillation elements (DSE) -- one element per
channel.
\item Continuous solid plates of a single crystal (SCP) -- by the size
of the photosensitive area of a multi-channel photoreceiver.
\item Plates of dispersed scintillator (DSP) -- made of tiny calibrated
particles of ZnSe(Te) crystals arranged into a monolayer and
optical epoxy adhesive UP4-20-3M [3].
\end{itemize}

Fabrication of plates composed of DSE at the step of
photosensitive elements in the array less than 0.8-0.5 $mm$ is
limited by technological difficulties. Therefore, with the aim of
looking for alternative ways to improve the resolution, we studied
properties of SCP and DSP. The plates were compared by two main
parameters -- light output and interference of the neighboring
channels.

\begin{figure}[ht]
\begin{center}
\includegraphics*[scale=1.0]{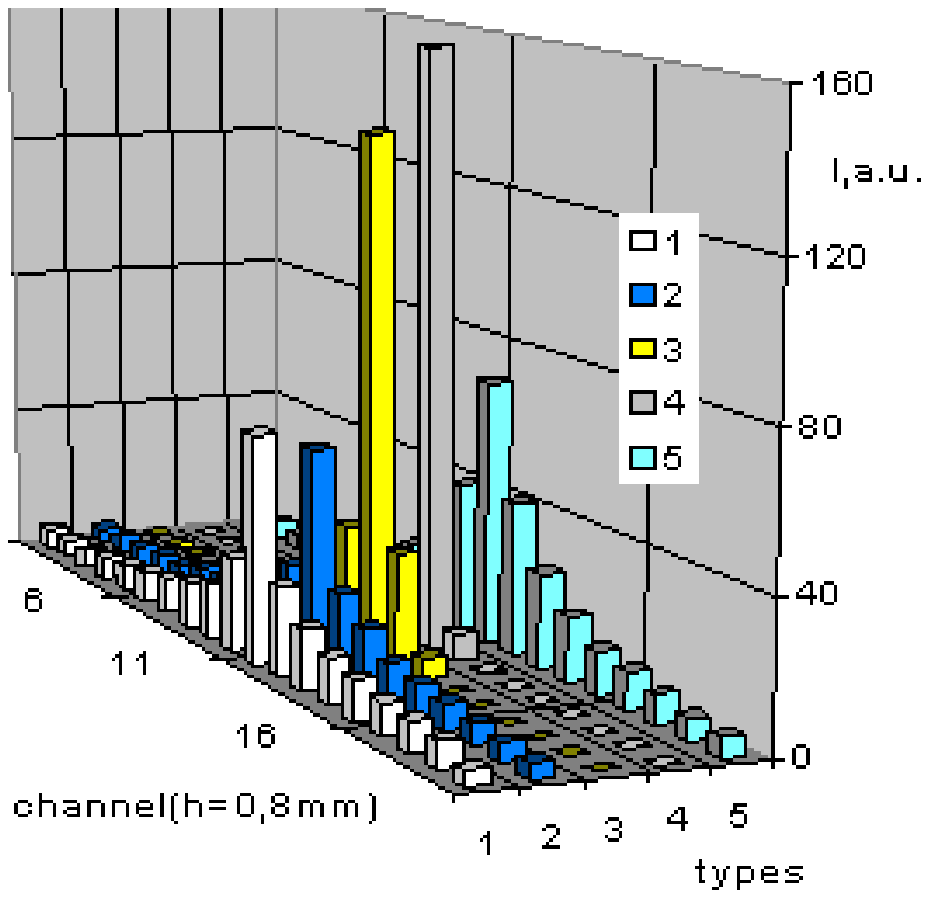}
\end{center}
\caption{Mutual interference of neighboring channels for different
scintillators and types of scintillation elements: 1 -- single
crystal plate ZnSe(Te) with $h=0.8 mm$; 2 -- single crystal plate
ZnSe(Te) with $h=0.6 mm$; 3 -- composite small-crystalline plate
ZnSe(Te) (grain size $0.4 mm$); 4 -- individual single elements
for each channel; 5 -- single crystal plate CsI(Tl) with $h=0.8
mm$ .}
\end{figure}

\begin{figure}[ht]
\begin{center}
\includegraphics*[scale=1.0]{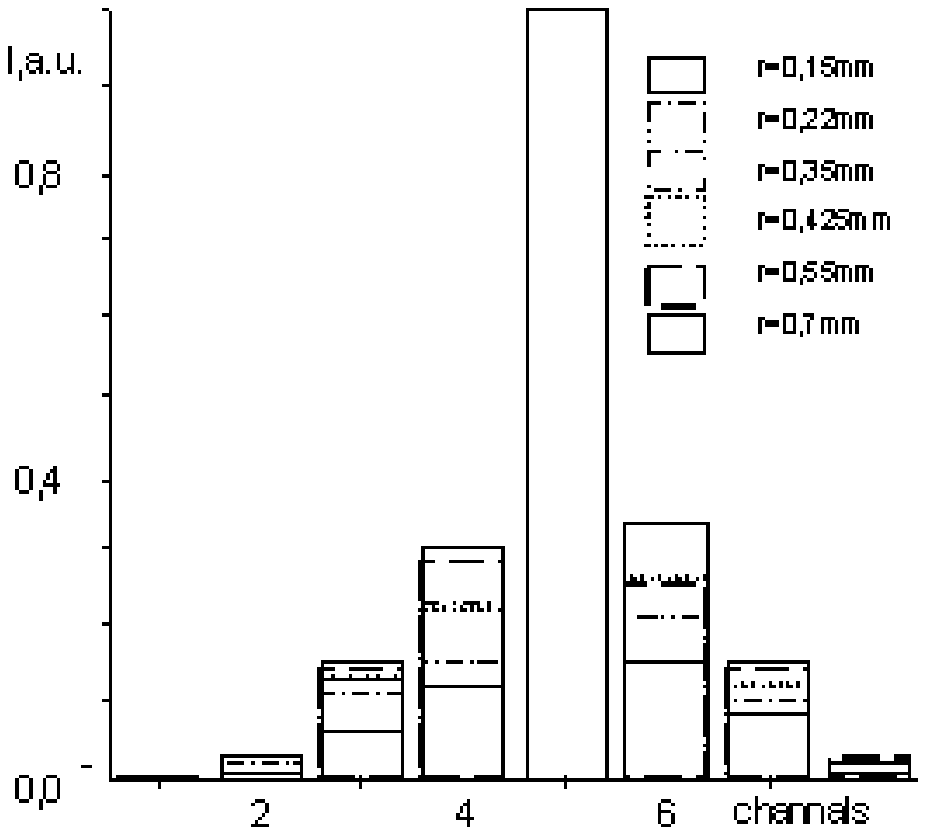}
\end{center}
\caption{Interference of neighboring channels in 16-element
detectors for a dispersed scintillation plate as function of
ZnSe(Te) scintillator grain size.}
\end{figure}

Comparative measurements have shown (Fig.2) that serious
competition to DSE plates can come only from DSP. Interference of
neighboring channels was studied for DSP prepared from ZnSe(Te)
grains of different size (Fig.3). It is obvious that with smaller
grain size the interference of neighboring channels is reduced.
However, the DSP light output is also dependent upon grain size
and has maximum at the grain size $\sim 0.5 mm$ (Fig.4). Thus,
using DSP, it is possible to make detector modules with step of
the photosensitive elements $0.2-0.5 mm$ for detection of X-ray
radiation in the energy range $20-80 keV$. Resolution of 1-2 line
pairs per $mm$ can be achieved.

The use of laws of geometrical optics, namely, placing the
inspected object much closer to the source of sharp focus than to
the detector array, allows one, using DSE with step 0.8 mm, to
achieve detecting abilities of 10-15 $\mu m$ (Fig.5, microchip).
Mathematical processing of the digitalized signal using a modern
software can substantially increase the informativity of the image
(Fig.6, the same chip).

The image quality is also determined by the choice of a
scintillator. Accounting for much higher, as compared with
CsI(Tl), quantum yield of ZnSe(Te) (by 20-30\% ), it has clear
advantages for low-energy detectors. With higher energies (Fig.7),
CsI(Tl) becomes preferable because of its high transparence.
However, if there are requirements for low afterglow level or high
radiation stability, ZnSe(Te) can be used up to the energies of
150-200 $keV$. At energies above $1 MeV$, superiority goes to
CdWO$_{4}$.

\begin{figure}[ht]
\begin{center}
\includegraphics*[scale=1.0]{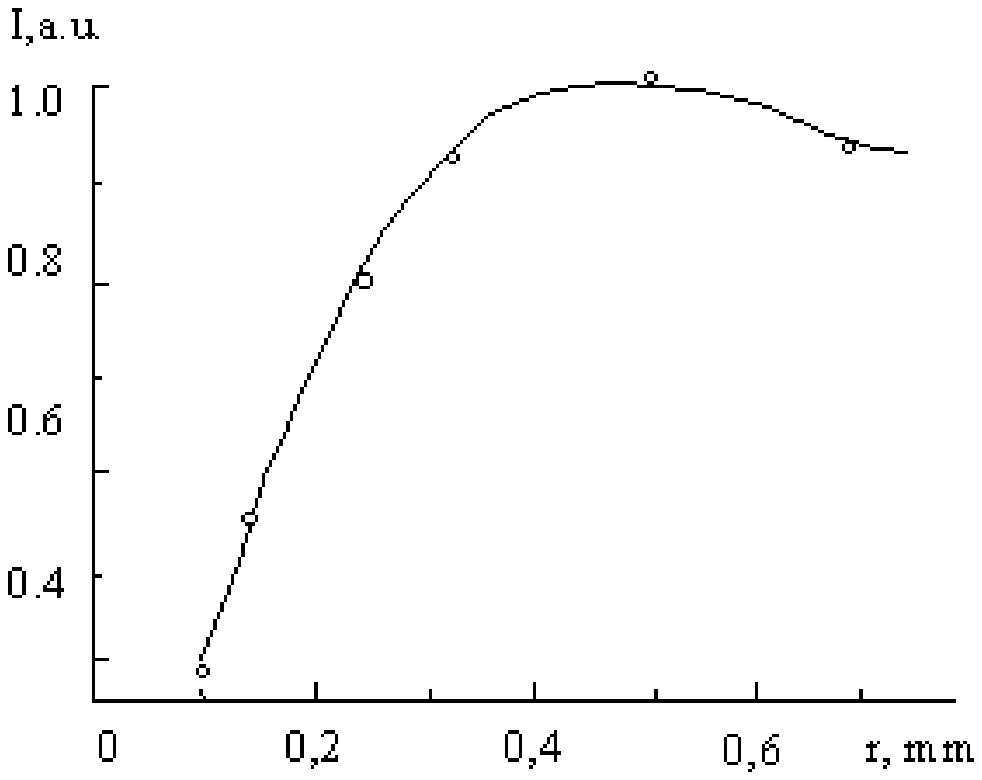}
\end{center}
\caption{Light output dependence upon grain size.}
\end{figure}

\begin{figure}[ht]
\begin{center}
\includegraphics*[scale=0.7]{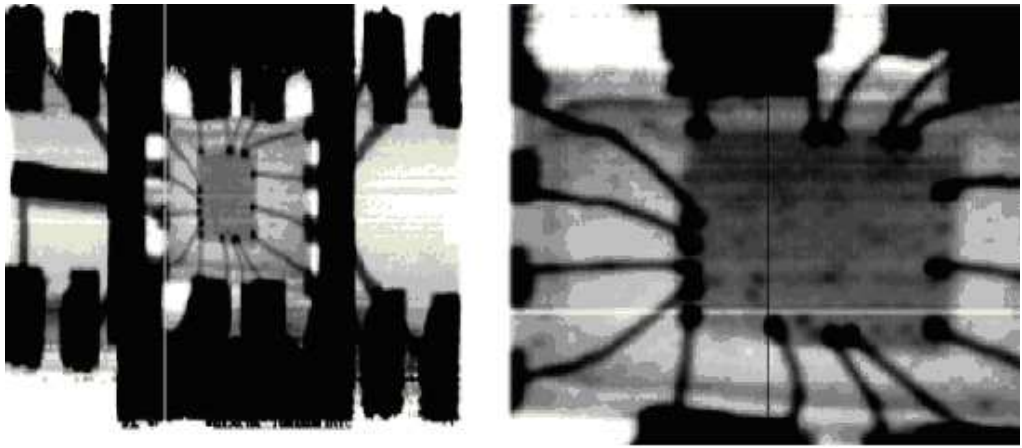}
\end{center}
\caption{Internal connection wire pattern from the semiconductor
crystal in a microchip.}
\end{figure}

\begin{figure}[ht]
\begin{center}
\includegraphics*[scale=1.0]{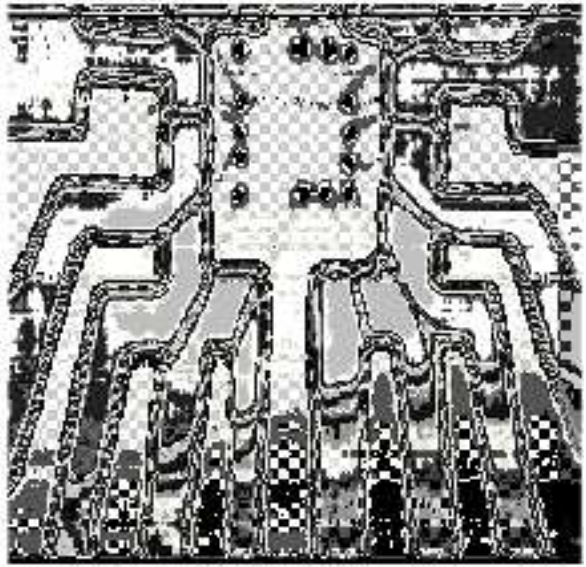}
\end{center}
\caption{Microchip in plastic tank.}
\end{figure}

\begin{figure}[ht]
\begin{center}
\includegraphics*[scale=0.80]{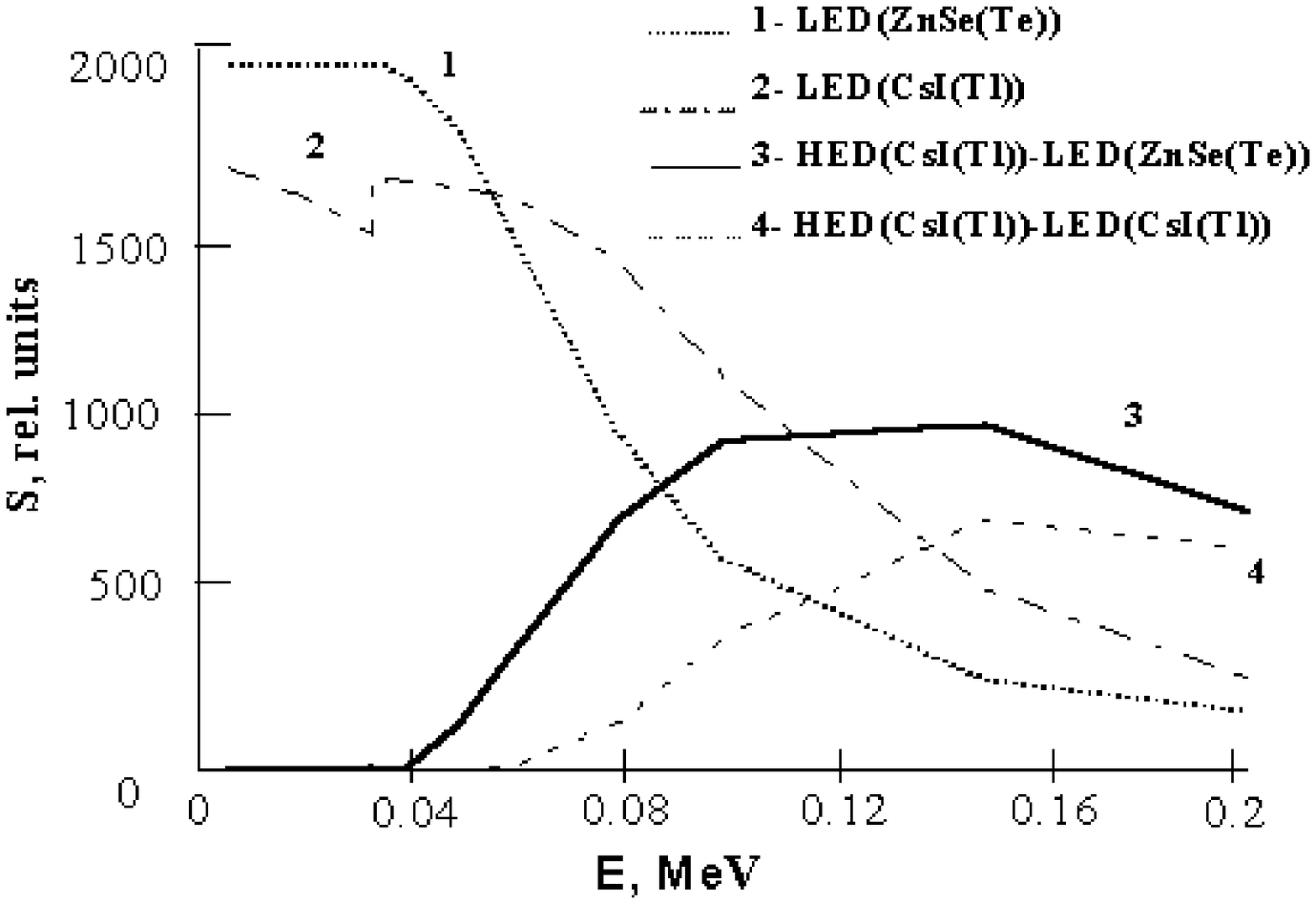}
\end{center}
\caption{Sensitivity of detectors using CsI(Tl) and ZnSe(Te) as
function of the radiation energy.}
\end{figure}

\section{Discussion}

Results of our studies have shown that in S-PD detectors for
digital radiography the most preferable scintillators for the high
energy region are CWO (0.5-1.0 $MeV$), CsI(Tl) (0.08-0.5 $MeV$),
while in the low-energy region (20-60 $keV$) ZnSe(Te) is
unchallenged.

Accounting for a trend in the modern digital radiography (both in
the inspection and medical instruments) to use two-energy detector
arrays, the combination of ZnSe(Te) and CsI(Tl)/CWO results in a
new quality.

Effective atomic number $Z$ of ZnSe is the same as of copper,
which is usually used as a filter of the high-energy array.
Therefore, if a detector with ZnSe(Te) as filter is placed before
the high-energy array, this simplifies the design and improves
technical characteristics of the detecting circuit as a whole [9].

A unique combination of properties characterizing the original
scintillator ZnSe(Te) -- high light output, fast response,
radiation stability, rather low effective atomic number together
with sufficiently high density -- makes this material the best
among known scintillators for the low-energy detector. Combination
of crystals ZnSe(Te)/CsI(Tl) in the two-energy detector array has
substantially improved the sensitivity of equipment designed for
detection of organic inclusions.

Theoretical estimates show that transition from the two-energy to
multi-energy detection system will allow distinguishing between
substances that differ in their effective atomic number
insignificantly. In the applied aspect, this means a possibility
to reliably (close to 100\% ) identify dangerous substances
(explosives, drugs) on the background of metal or organics.

A logical way to better resolution in digital radiography is
decreasing the aperture of a unit detector. Experimental data
obtained in this work show that with the existing standard
production technology of the arrays and the S-PD detector as a
whole it would be very difficult to make them with step less than
0.8-1 mm. At the same time, if there is a discrete amplifier for
each channel, requirements to the photodiode-preamplifier system
become more complex. In this relationship, for detector arrays of
step 0.05-0.8 $mm$ that have been studied in the present work, it
is desirable to pass over to a solid scintillation plate (single
crystal or dispersed scintillator) and, in parallel, to replace
the photodiode with a photoreceiving unit with an
amplifier-commutator at the back side. This allows the use of just
one amplifier channel after a photoline of 32-1024 channels of the
photoreceiving device.

Limitations imposed upon detectors "scintillator plate --
photoreceiving device" are, from the one side, the requirement
for the plate to be thin as to decrease interference of the
neighboring channels, which limits the energy range of the
detector. From the other side, low radiation stability of modern
PRD requires their shielding from radiation, i.e., their use at
low dose loads, which is just the conditions used in inspection
equipment and medicine. For systems of technical digital
radiography and tomography, the preferable solution is discrete
arrays S-PD, with ZnSe(Te) and CWO crystals of high radiation
stability being the recommended scintillators, and photodiodes as
p-n structures, desirably in the photogalvanic mode. Using such
structures, after amplification under the laws of geometrical
optics, we have obtained, in the tomographic mode with transition
to 3D image, sufficiently high spatial resolution (Fig.8).

\begin{figure}[ht]
\begin{center}
\includegraphics*[scale=1.0]{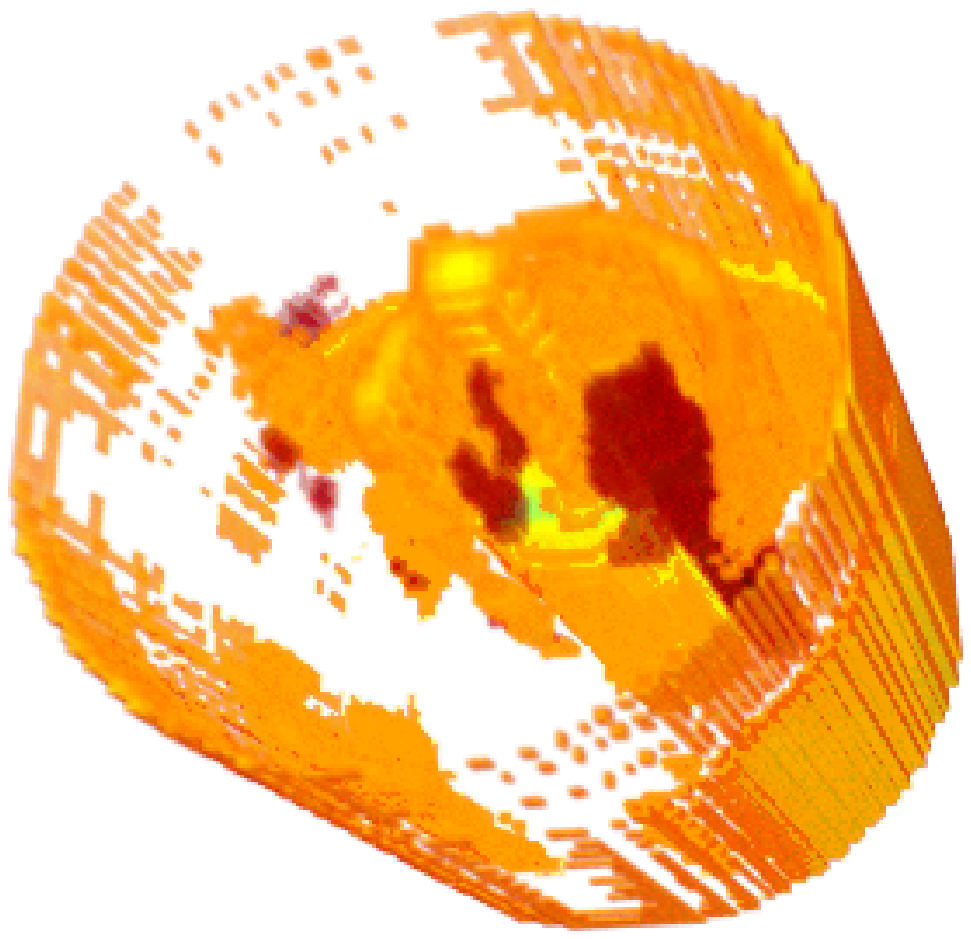}
\end{center}
\caption{Tomographic image of the lighter friction head.}
\end{figure}

\bigskip

\section{Reference}

\noindent [1] R.M.~Harrison, "Digital radiography -- a review of
detector design", \textit{Nucl. Instr. and Meth}. \textbf{A310},
24-34 (1991).

\noindent [2] V.D.~Ryzhikov, N.G.~Starzhinskiy,
L.P.~Gal'chinetskii, D.N.~Kozin, V.P.~Sokhin, A.D.~Opolonin,
V.M.~Svishch, E.K.~Lisetskaya, "Scintillator-photodiode detecting
systems for two-level X-ray inspection systems", 15th World
Conference on Non-Destructive Testing, Rome (Italy), 15-21
October, 2000, Abstracts Book, p. 466.

\noindent [3] S.V.~Naydenov, V.D.~Ryzhikov, "Determining Chemical
Compositions by Method of Multi-Energy Radiography",
\textit{Technical Physics Letters} \textbf{28}, \# 5, 357-360
(2002).

\noindent [4] Heimann, Prospects, 2002,
\textbf{http://www.heimannsystems.com}.

\noindent [5] The X-Ray Introscopy System "Poliscan-4", Prospects,
{stcri@isc.kharkov.com}.

\noindent [6] L.V.~Atroshchenko, B.V.~Grinev, V.D.~Ryzhikov et al.
(ed. by V.D.~ Ryzhikov), \textit{Scintillator crystals and
detectors of ionizing radiation on their base}, Kiev, Naukova
Dumka, 1988 (in Russian).

\noindent [7] V.G.~Volkov, B.V.~Grinev, V.D.~Ryzhikov et. al.,
"Small-crystalline detectors of ionizing radiation on the basis of
ZnSe(Te)", \textit{Pribory i Tekhnika Eksperimenta} (Moscow),
No.~6, p.~37-42 (1999).

\noindent [8] Detection system for X-ray introscopy, Patent of
Ukraine, No.~2001053401 from 21.05.2001.

\end{document}